 \documentclass[final,5p,times,twocolumn]{elsarticle}

\usepackage{graphicx}

\usepackage{amssymb}

\journal{Nuclear Instruments and Methods in Physics Research Section B}

\begin{document}

\begin{frontmatter}



\title{Role of exchange interaction in self-consistent calculations of endohedral fullerenes}


\author{Alexey V. Verkhovtsev}
\ead{verkhovtsev@physics.spbstu.ru}
\author{Roman G. Polozkov}
\author{Vadim K. Ivanov}
\address{St. Petersburg State Polytechnic University, Politekhnicheskaya ul. 29, 195251 St. Petersburg, Russia}

\author{Andrei V. Korol}
\author{Andrey V. Solov'yov}
\address{Frankfurt Institute for Advanced Studies, Ruth-Moufang-Str. 1, 60438 Frankfurt am Main, Germany}

\begin{abstract}
Results of the self-consistent calculation of electronic structure of endohedral fullerene Ar@C$_{60}$ within the
Hartree-Fock and the local density approximations are presented. Hartree-Fock approximation is used for the self-consistent description for the first time. It is shown that the accurate account of the exchange interaction between all electrons of the compound leads to the significant modification of the atomic valent shell which causes the noticeable charge redistribution inside the endohedral compound.
\end{abstract}


\end{frontmatter}

\section{Introduction}

Endohedral fullerenes represent a class of compounds consisted of atoms or small molecules confined inside the fullerene cage. Due to the confinement of the encaged atom endofullerenes have unique properties and a wide range of potential applications. For instance, they are used in drug delivery applications \cite{DrugDelivery}, doping of endohedral fullerenes allows one to improve power conversion efficiency of the photovoltaic devices \cite{Ross_2009_NatureMaterials.8.208}. Quantum dot devices based on endohedral fullerenes encapsulated in single-walled carbon nanotubes are reported to be created \cite{Lee_2002_Nature.415.1005} and proposed for quantum-information processing \cite{Yang_2010_PhysRevA.81.032303}.

Besides the potential applications the investigation of fundamental properties of endohedral fullerenes is also of a great current interest. Since the first experimental observation of endohedral fullerenes \cite{Heath_1985_JAmChemSoc.107.7779} a large number of theoretical investigations have been performed on the photoionization process of the compounds within different frameworks (see, e.g., review \cite{Dolmatov_2009_AdvQuantChem_review} and references therein). On the contrary, the experimental studies are still not numerous what is caused by difficulties in producing sufficient amounts of purified compounds for the gas phase experiments. Up to now series of experiments have been performed on the neutral and charged endometallofullerenes \cite{Mitsuke_Ce@C82_2005_JChemPhys.122.064304, Mueller_Ce@C82+_2008_PhysRevLett.101.133001, Katayanagi_Pr@C82_2008_JQuantSpectrRadTrans.109.1590} and on noble gas endohedral fullerenes, Ar@C$_{60}$ \cite{Morscher_2010_PhysRevA, Takagi_Dragoe_2011_PhysChemChemPhys.13.9609} and Xe@C$_{60}^+$ \cite{Mueller_Xe@C60_2010_PhysRevLett.105.213001} as well.

In this paper we present the results of the first self-consistent description of electronic structure of endohedral fullerene Ar@C$_{60}$ within the Hartree-Fock (HF) approximation. From atomic physics it is known \cite{Amusia_AtPhotoeff} that an accurate description of the photoionization process of many-electron systems is obtained within the Random Phase Approximation with Exchange (RPAE) which is based on application of the HF approximation for calculation of the ground and excited states of the system. It was proven by case studies of a variety of isolated atoms and their ions \cite{Ivanov_review_1999_JPhysB.32.R67} and of metallic clusters as well \cite{Solovyov_2001_Clusters}.

Considering Ar@C$_{60}$ within the local density approximation (LDA) in \cite{Madjet_2007_PhysRevLett.99.243003} the hypothesis of hybridization of the valent 3p shell of the confined Ar atom was discussed for the first time. In the present paper we describe the hybridization phenomena in another way and find out the correlated effect caused by the hybridization. Application of the HF approximation allows one to take into account the exchange interaction between all considered electrons of the system more accurately in comparison with the LDA.

\section{Model description}

In a number of papers it was supposed that in noble gas endohedral fullerenes A@C$_{60}$ a confined atom A is located in the center of the fullerene molecule \cite{Albert_2007_IntJQuantChem.107.3061, Pyykko_2007_PhysChemChemPhys.9.2954, Pyykko_2010_PhysChemChemPhys.12.6187} vibrating due to finite temperature \cite{Korol_2010_JPhysB.43.201004}. In the present paper we assume that in case of Ar@C$_{60}$ these vibrations are insignificant at room temperature and consider Ar atom located in the center of the fullerene cage. Thereby all the electrons of the endohedral system are moving in a spherically symmetric central field and one can construct the electronic configuration described by the unique set of quantum numbers $\{n,l\}$ where $n$ and $l$ are the principal and the orbital quantum numbers correspondingly.

In the present paper the following scheme of a shell filling is proposed. Electronic configuration of an isolated Ar atom is 1s$^2$2s$^2$2p$^6$3s$^2$3p$^6$. Using the atomic classification, the configuration of 240 delocalized electrons in the pristine C$_{60}$ can be written as:
\begin{eqnarray}
\textrm{1s}^2\textrm{2p}^6\textrm{3d}^{10}\textrm{4f}^{14}\textrm{5g}^{18}\textrm{6h}^{22}\textrm{7i}^{26}
\textrm{8k}^{30}\textrm{9l}^{34}\textrm{10m}^{18} \nonumber \\
\textrm{2s}^2\textrm{3p}^6\textrm{4d}^{10}\textrm{5f}^{14}\textrm{6g}^{18}\textrm{7h}^{10}, \label{eq1}
\end{eqnarray}
where the first 10 shells (1s \dots 10m) are nodeless and the last 6 shells (2s \dots 7h) have a single radial node. Since it is commonly acknowledged that fullerenes are formed from the fragments of the planar graphite sheets, it is natural to match $\sigma$- and $\pi$-orbitals of the graphite to the nodeless and the single-node wave functions of a fullerene, respectively \cite{Martins_1991_ChemPhysLett.180.457}. Different graphite sheets are connected by weak $\pi$-bonds whereas the carbon atoms of a same sheet are connected by $\sigma$-bonds. In the fullerene the nodeless $\sigma$-orbitals are localized at the radius of the cage while the single-node $\pi$-orbitals, on the contrary, are oriented perpendicularly to the surface of the fullerene. The ratio of $\sigma$- and $\pi$-orbitals in C$_{60}$ should be equal to $3 : 1$ due to the sp$^2$-hybridization of carbon orbitals \cite{Haddon_1986_ChemPhysLett.125.459}.

All 258 electrons of the endohedral compound Ar@C$_{60}$ are described within a framework of the united electronic configuration. The filling scheme is as follows. Double-nodal 3s$^2$ shell is added to the configuration of the pristine C$_{60}$ (\ref{eq1}). The latter 16 electrons are distributed on the non-closed 10m and 7h shells so that the number of electrons located on $\sigma$- and $\pi$-orbitals is conserved. As a result the electronic configuration of the endohedral compound is written as
\begin{eqnarray}
\textrm{1s}^2\textrm{2s}^2\textrm{2p}^6\textrm{3s}^2\textrm{3p}^6\textrm{3d}^{10}\textrm{4f}^{14}\textrm{5g}^{18}
\textrm{6h}^{22}\textrm{7i}^{26}\textrm{8k}^{30}\textrm{9l}^{34}\textrm{10m}^{26} \nonumber \\
\textrm{4d}^{10}\textrm{5f}^{14}\textrm{6g}^{18}\textrm{7h}^{18}. \label{eq2}
\end{eqnarray}

To define single-electron energies and wave functions the system of self-consistent equations is solved for all electrons of the compound. The potential of positive ions, $U$, is included in the equations and combines the potential of the fullerene core and the Coulomb potential due to the nucleus $Z$ of the confined atom ($Z = 18$ in case of Ar), $U = U_{\textrm{core}} - Z/r$. Electronic system is treated in the HF and the LDA approximations.

The fullerene core of 60 fourfold charged carbon ions C$^{4+}$ is described via jellium model as a positively charged spherical layer of a finite $\Delta R$ thickness.
The average radius $R$ and $\Delta R$ are the adjustable parameters of the potential. In the current calculations $R$ stands for the radius of C$_{60}$ which is equal to $3.54$ \AA \ and $\Delta R$ is taken equal to $1.5$ \AA \ \cite{Ruedel_2002_PhysRevLett.89.125503}.

\section{Results}

The results of the self-consistent calculation of electronic structure of Ar@C$_{60}$ are presented below. Figure \ref{figure1} represents the radial potential of the compound calculated within the HF and the LDA frameworks, the 3p and 4d wave functions of the compound are presented in Figures \ref{figure2} and \ref{figure3}. The local part of the HF and the LDA radial self-consistent potentials are mapped in Figure \ref{figure1} by the triangles and the filled circles, respectively. As far as the system of equations is solved with different potentials within the HF approximation and the LDA, wave functions obtained within these approximations have a remarkable difference. One can see that accounting for the exact non-local exchange interaction within the HF approximation reveals significant modification of the 3p and 4d wave functions (the line with triangles in Figures \ref{figure2} and \ref{figure3}). On the contrary, accounting for the local exchange interaction within the LDA doesn't lead to the such modification of the wave functions (the filled-circled line in Figures \ref{figure2} and \ref{figure3}).

\begin{figure} [h]
\centering
\includegraphics[scale=0.32,clip]{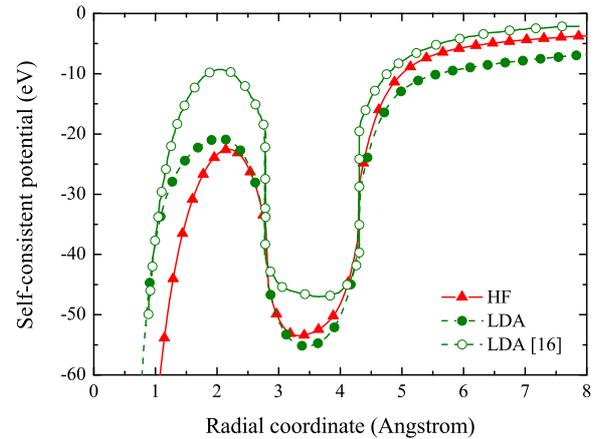}
\caption {Radial self-consistent potential of Ar@C$_{60}$ calculated within the HF approximation (line with triangles) and the LDA (filled-circled line). Results of the LDA calculations performed by Madjet et al. \cite{Madjet_2007_PhysRevLett.99.243003} are also presented (open-circled line).}\label{figure1}
\end{figure}

In Ref. \cite{Madjet_2007_PhysRevLett.99.243003} an assumption on the hybridization of the argon 3p shell in Ar@C$_{60}$ was formulated. In the cited paper the system of the LDA equations was solved for the superposition of atomic state and the fullerene state (see \cite{PhysRevA.81.013202} for the details). The hybridized 3p state was defined as a superposition of the 3p shell of a free argon atom and the 3p shell in pristine C$_{60}$. Due to the mixing of the atomic and fullerene wave functions the additional node appears in the hybridized 3p function (the open-circled line in Figure \ref{figure2}). The radial potential of the mentioned calculation is presented in Figure \ref{figure1} by the open-circled line.

\begin{figure} [h]
\centering
\includegraphics[scale=0.32,clip]{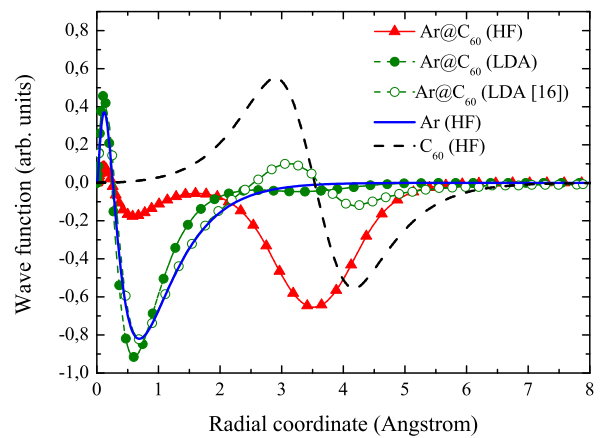}
\caption {The 3p wave function in free Ar atom (solid line), in pristine C$_{60}$ (dashed line) and in Ar@C$_{60}$ calculated within the HF approximation (line with triangles) and the LDA (filled-circled line). Results of the LDA calculations performed by Madjet et al. \cite{Madjet_2007_PhysRevLett.99.243003} are also presented (open-circled line).}\label{figure2}
\end{figure}

\begin{figure} [h]
\centering
\includegraphics[scale=0.32,clip]{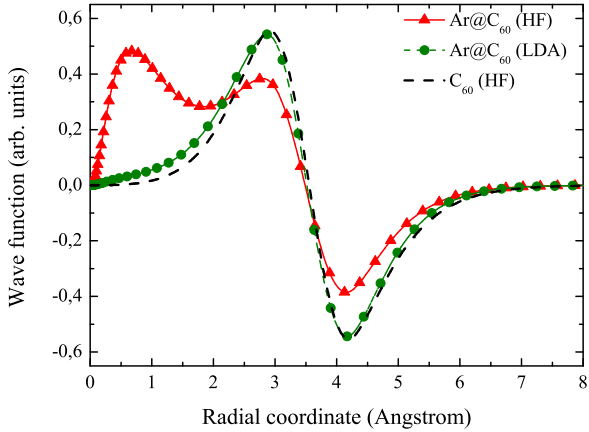}
\caption {The 4d wave function in pristine C$_{60}$ (dashed line) and in Ar@C$_{60}$ calculated within the HF approximation (line with triangles) and the LDA (filled-circled line).}\label{figure3}
\end{figure}

As mentioned above, in the present work we introduce the unified electronic configuration. In spite of the remarkable redistribution of the electronic density the 3p wave function has a single radial node as opposed to the framework used in Ref. \cite{Madjet_2007_PhysRevLett.99.243003}. Let us note, however, that the symmetry of the wave function is significantly changed. The 3p electronic density in the pristine C$_{60}$ has a minimum at the radius of the fullerene, so it corresponds to the $\pi$-orbital (the dashed line in Figure \ref{figure2}). In the case of Ar@C$_{60}$ the 3p shell becomes a $\sigma$-type orbital with the maximum distribution of the electronic density at the radius of the fullerene (the line with triangles). The change of the symmetry and the redistribution of electronic density allows one to draw a conclusion on the significant hybridization of the 3p shell within the HF approximation. This effect disappears if the compound is treated within the LDA due to account of the local exchange interaction between the electrons of the confined atom and the delocalized electrons of the fullerene. This result generally correlates to the result previously obtained within the LDA in \cite{Madjet_2007_PhysRevLett.99.243003} (see the open-circled line in Figure \ref{figure2}).

\begin{figure} [h]
\centering
\includegraphics[scale=0.32,clip]{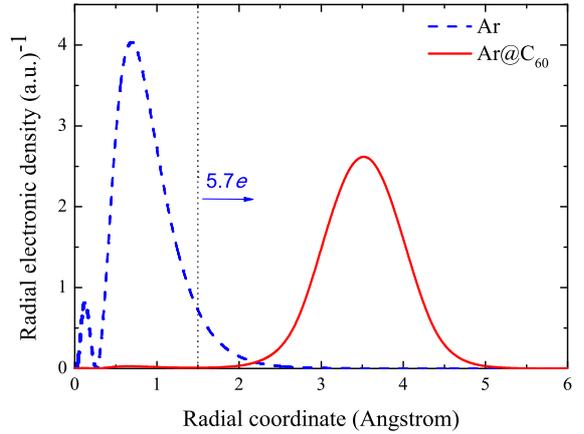}
\caption {Redistribution of the electronic density of the 3p shell in Ar@C$_{60}$. }\label{figure4}
\end{figure}

\begin{figure} [h]
\centering
\includegraphics[scale=0.32,clip]{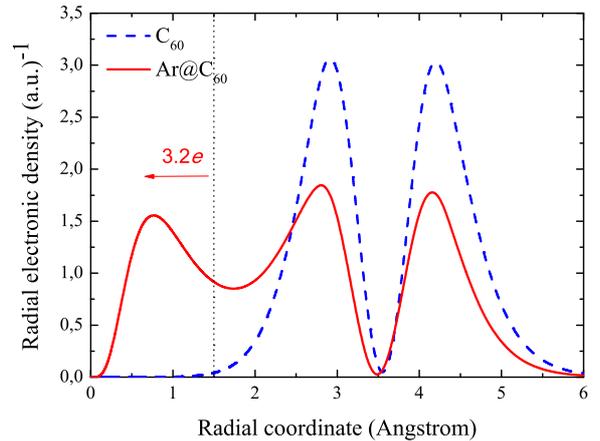}
\caption {Redistribution of the electronic density of the 4d shell in Ar@C$_{60}$. }\label{figure5}
\end{figure}

For the considered electronic configuration (\ref{eq2}) the account of the non-local exchange interaction within the HF approximation leads to the remarkable charge redistribution due to the modification of the 3p and 4d shells (see Figures \ref{figure4} and \ref{figure5}). As it can be seen in Figure \ref{figure4} the most part of the electronic density of the 3p shell of free Ar is located within the sphere of 1.5 \AA. But in the endohedral compound the partial charge of the 3p$^6$ shell equal to 5.7$e$ (where $e$ stands for the elementary charge) is pressed out of this sphere of 1.5 \AA \ to the vicinity of the fullerene's cage. In contrary, the partial charge of the 4d$^{10}$ shell which is equal to 3.2$e$ is transferred from the vicinity of the fullerene's cage to the vicinity of the confined atom.

\section{Conclusion}

To conclude, the self-consistent Hartree-Fock description of the endohedral fullerene Ar@C$_{60}$ was carried out for the first time. Electrons of the encaged atom and the delocalized electrons of C$_{60}$ were considered simultaneously by introduction of the united electronic configuration. Having considered the configuration described above it was shown that the account of the non-local exchange interaction within the Hartree-Fock approximation leads to the significant modification of the 3p and 4d shells as opposed to the local exchange interaction within the local density approximation.
As a result of the modification the redistribution of the electronic density of the 3p and 4d shells appears and causes the accumulation of the additional positive charge in the vicinity of the encaged atom and the additional negative charge on the fullerene core.

To accomplish in full the analysis of the phenomena of hybridization and shells modification in Ar@C$_{60}$ one has to investigate electronic configurations other than in (\ref{eq2}). This work is currently in progress and the results will be published elsewhere.

\section*{Acknowledgements}

A.V.V. would like to thank the Deutscher Akademischer Austausch Dienst (DAAD) for the financial support.



\section*{References}




\begin{thebibliography}{50}

\bibitem{DrugDelivery}
J.B. Melanko, M.E. Pearce, A.K. Salem, Nanotubes, Nanorods, Nanofibers, and Fullerenes for Nanoscale Drug Delivery, in: M.M. de Villiers, P. Aramwit, G.S. Kwon (eds.), Nanotechnology in Drug Delivery, Springer Science+Business Media, New York, 2009, p. 105.

\bibitem{Ross_2009_NatureMaterials.8.208}
R.B. Ross, C.M. Cardona, D.M. Guldi, S.G. Sankaranarayanan, M.O. Reese, N. Kopidakis, J. Peet, B. Walker, G.C. Bazan, E. Van Keuren, B.C. Holloway, M. Drees, Nature Materials 8 (2009) 208.

\bibitem{Lee_2002_Nature.415.1005}
J. Lee, H. Kim, S.-J. Kahng, G. Kim, Y.-W. Son, J. Ihm, H. Kato, Z.W. Wang, T. Okazaki, H. Shinohara, Y. Kuk, Nature 415 (2002) 1005.

\bibitem{Yang_2010_PhysRevA.81.032303}
W.L. Yang, Z.Y. Xu, H. Wei, M. Feng, D. Suter, Phys. Rev. A 81 (2010) 032303.

\bibitem{Heath_1985_JAmChemSoc.107.7779}
J.R. Heath, S.C. O'Brien, Q. Zhang, Y. Liu, R.F. Curl, H.W. Kroto, F.K. Tittel, R.E. Smalley, J. Am. Chem. Soc. 107 (1985) 7779.

\bibitem{Dolmatov_2009_AdvQuantChem_review}
V.K. Dolmatov, Theory of Confined Quantum Systems, in: J.R. Sabin, E. Br\"{a}ndas (eds.), Advances in Quantum Chemistry, Academic Press, New York, vol. 58, 2009, p. 13.

\bibitem{Mitsuke_Ce@C82_2005_JChemPhys.122.064304}
K. Mitsuke, T. Mori, J. Kou, Y. Haruyama, Y. Kubozono, J. Chem. Phys. 122 (2005) 064304.

\bibitem{Mueller_Ce@C82+_2008_PhysRevLett.101.133001}
A. M\"uller, S. Schippers, M. Habibi, D. Esteves, J.C. Wang, R.A. Phaneuf, A.L.D. Kilcoyne, A. Aguilar, L. Dunsch, Phys. Rev. Lett. 101 (2008) 133001.

\bibitem{Katayanagi_Pr@C82_2008_JQuantSpectrRadTrans.109.1590}
H. Katayanagi, B.P. Kafle, J. Kou, T. Mori, K. Mitsuke, Y. Takabayashi, E. Kuwahara, Y. Kubozono, J. Quant. Spectr. Rad. Transfer 109 (2008) 1590.

\bibitem{Morscher_2010_PhysRevA}
M. Morscher, A. Seitsonen, S. Ito, H. Takagi, N. Dragoe, T. Greber, Phys. Rev. A 82 (2010) 051201(R).

\bibitem{Takagi_Dragoe_2011_PhysChemChemPhys.13.9609}
F. Cimpoesu, S. Ito, H. Shimotani, H. Takagi, N. Dragoe, Phys. Chem. Chem. Phys. 13 (2011) 9609.

\bibitem{Mueller_Xe@C60_2010_PhysRevLett.105.213001}
A.L.D. Kilcoyne, A. Aguilar, A. M\"uller, S. Schippers, C. Cisneros, G. Alna'Washi, N.B. Aryal, K.K. Baral, D.A. Esteves, C. M. Thomas, R.A. Phaneuf, Phys. Rev. Lett. 105 (2010) 213001.

\bibitem{Amusia_AtPhotoeff}
M.Ya. Amusia, Atomic Photoeffect, Springer-Verlag New York, LLC, 1990.

\bibitem{Ivanov_review_1999_JPhysB.32.R67}
V.K. Ivanov, J. Phys. B 32 (1999) R67.

\bibitem{Solovyov_2001_Clusters}
A.V. Solov'yov, Electron Scattering on Metal Clusters and Fullerenes, in: C. Guet, P. Hobza, F. Spiegelman, F. David (eds.), Atomic Clusters and Nanoparticles, EDP Sciences, Springer-Verlag, vol. 73, 2001, pp. 401-435.

\bibitem{Madjet_2007_PhysRevLett.99.243003}
M.E. Madjet, H.S. Chakraborty, S.T. Manson, Phys. Rev. Lett. 99 (2007) 243003.

\bibitem{Albert_2007_IntJQuantChem.107.3061}
V.A. Albert, J.R. Sabin, F.E. Harris, Int. J. Quant. Chem. 107 (2007) 3061.

\bibitem{Pyykko_2007_PhysChemChemPhys.9.2954}
P. Pyykk\"o, C. Wang, M. Straka, J. Vaara, Phys. Chem. Chem. Phys. 9 (2007) 2954.

\bibitem{Pyykko_2010_PhysChemChemPhys.12.6187}
C. Wang, M. Straka, P. Pyykk\"o, Phys. Chem. Chem. Phys. 12 (2010) 6187.

\bibitem{Korol_2010_JPhysB.43.201004}
A.V. Korol, A.V. Solov'yov, J. Phys. B 43 (2010) 201004.

\bibitem{Martins_1991_ChemPhysLett.180.457}
J.L. Martins, N. Troullier, J.H. Weaver, Chem. Phys. Lett. 180 (1991) 457.

\bibitem{Haddon_1986_ChemPhysLett.125.459}
R.C. Haddon, L.E. Brus, K. Raghavachari, Chem. Phys. Lett. 125 (1986) 459.

\bibitem{Ruedel_2002_PhysRevLett.89.125503}
A. R\"udel, R. Hentges, U. Becker, H.S. Chakraborty, M.E. Madjet, J.-M. Rost, Phys. Rev. Lett. 89 (2002) 125503.

\bibitem{PhysRevA.81.013202}
A. Potter, M.A. McCune, R. De, M.E. Madjet, H.S. Chakraborty, Phys. Rev. A 82 (2010) 033201.

\end{thebibliography}



\end{document}